\documentclass[aip,apl,reprint]{revtex4-1}

\usepackage{graphicx}

\begin{document}

\title{Direct detection of magnon spin transport by the inverse spin Hall effect}

\author{A. V. Chumak}

\email{chumak@physik.uni-kl.de}

\author{A. A. Serga}

\author{M. B. Jungfleisch}

\author{R. Neb}

\affiliation{Fachbereich Physik and Forschungszentrum OPTIMAS, Technische Universit\"at Kaiserslautern, 67663
Kaiserslautern, Germany}

\author{D. A. Bozhko}

\affiliation{Faculty of Radiophysics, Taras Shevchenko National University of Kyiv, Kyiv, Ukraine}

\author{V. S. Tiberkevich}
\affiliation{Department of Physics, Oakland University, Rochester, MI 48309, USA}

\author{B.~Hillebrands}

\affiliation{Fachbereich Physik and Forschungszentrum OPTIMAS, Technische Universit\"at Kaiserslautern, 67663
Kaiserslautern, Germany}

\date{\today}

\begin{abstract}
Conversion of traveling magnons into an electron carried spin current is demonstrated in a time resolved experiment using a spatially separated inductive spin-wave source and an inverse spin Hall effect (ISHE) detector. A short spin-wave packet is excited in a yttrium-iron garnet (YIG) waveguide by a microwave signal and is detected at a distance of 3~mm by an attached Pt layer as a delayed ISHE voltage pulse. The delay in the detection appears due to the finite spin-wave group velocity and proves the magnon spin transport. The experiment suggests utilization of spin waves for the information transfer over macroscopic distances in spintronic devices and circuits.
\end{abstract}

\maketitle

Spin pumping (SP)\cite{tserkovnyak} and the inverse spin Hall effect (ISHE)\cite{saitoh-2006} are key mechanisms allowing conversion of magnons -- quanta of spin waves (SW)~\cite{Gurevich} excited in a ferromagnetic media into electron carried spin current and, consecutively, charge current in the attached nonmagnetic metal layer. The use of spin waves in spintronics allows the transfer of a spin angular momentum over macroscopic distances since the SW free path\cite{YIG-magnonics} is normally orders of magnitude larger than the spin-diffusion length in metals.\cite{bass-2007} Furthermore, a spin wave itself can be used for information processing (see for example Refs.~\cite{Realization-SW-logic, Wang-logic, chumak-NatComm, YIG-magnonics}).

At the same time, most of the works in the field deal with non-distant transducers and study standing SW modes,\cite{saitoh-2006, costache-PRL, ando-APL} practically non-propagating dipolar-exchange\cite{jungfleisch-APL} and short-wavelength exchange spin waves.\cite{sandweg-PRL, dzyapko-APL} Even the coupling between two closely placed Pt layers in Ref.~\cite{kajiwara2010} can be understood as a result of the excitation of a standing SW mode rather than of a traveling wave. Thus, no direct evidence on SP-ISHE detection of the magnon transport has been presented to date.

In this Letter, we use time-resolved measurements to demonstrate magnon spin transport between the spatially separated inductive spin-wave source and the SP-ISHE detector. The role of the traveling magnons is confirmed by a delay in the detection of the ISHE voltage pulse  associating with the SW packet propagation time.

\begin{figure}[t]
\includegraphics[width=0.95\columnwidth]{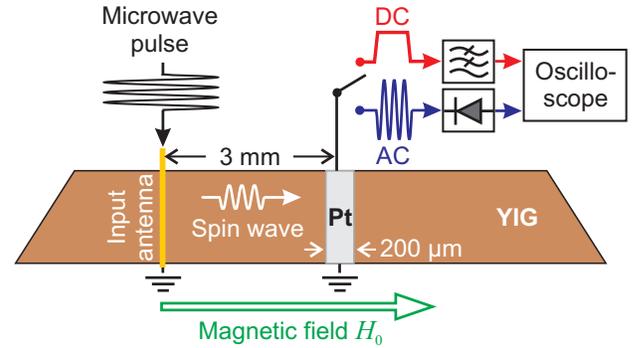}
\caption{\label{fig1} (Color online) Schematic illustration of the experimental setup: Spin-wave packet is excited in
the YIG waveguide using a microstrip antenna and detected at the 3~mm separated Pt stripe as electromagnetically
induced AC and ISHE voltage DC signals.}
\end{figure}

The structure under investigation is schematically illustrated in Fig.~\ref{fig1}. It comprises a 2.1~{$\mu$m} thick YIG waveguide ($19 \times 3$~mm$^{2}$) with a 10~nm thick ($0.2 \times 3$~mm$^{2}$, 374~Ohm) Pt strip deposited on the top.\cite{comment-1} The strip has been chosen to be sufficiently narrow to minimize distortions of spin waves due to their reflection and absorption. The YIG waveguide is magnetized along its long axis by an external bias magnetic field of $H_0 = 1754$~Oe providing conditions for the propagation of a backward volume magnetostatic wave (BVMSW).\cite{Gurevich} The BVMSW is excited by a microwave Oersted field of a 50~{$\mu$m}-wide Cu microstrip antenna placed at a distance of 3~mm from the Pt strip. The transmitted wave is detected at the Pt layer in two different ways (see Fig.~\ref{fig1}): (1) as an inductively excited AC microwave signal (like for the case of a conventional microstrip antenna) or (2) as an ISHE DC voltage. Both signals are measured using the same circuit comprising of a voltage preamplifier and an oscilloscope, but in the first case a microwave diode is utilized to rectify the signal (the name ``AC'' to denote the envelope of this signal), while for the DC measurements the diode is substituted by a low-pass filter.

\begin{figure}[b]
\includegraphics[width=0.95\columnwidth]{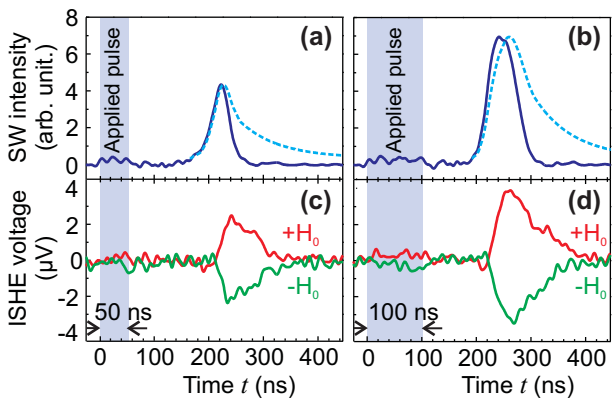}
\caption{\label{fig2} (Color online) Temporal evolution of the spin-wave intensity and the ISHE voltage for different field polarities. The dependencies are measured for the input signal durations $\tau_\mathrm{s} = 50$~ns (panels (a) and (c)), and 100~ns (panels (b) and (d)), magnetic field $H_0 = 1754$~Oe, applied power $P_\mathrm{s} = 10$~mV. Dashed lines in (a) and (b) show normalized total magnon densities $n_\mathrm{sum}$ calculated according to Eq.~(\ref{eq2}).}
\end{figure}

The experiment was performed in the following fashion: At time $t = 0$~ns the microwave pulse having a carrier frequency of $f_\mathrm{s} = 7$~GHz and the duration $\tau_\mathrm{s}$ in the range from 30~ns to 1~$\mu$s is applied to the input microstrip antenna. This results in the excitation of a travelling wave packet which propagates towards the Pt strip.\cite{YIG-magnonics} After a delay of approximately 200~ns determined by the SW group velocity the transmitted SW packet is detected as an AC signal at the Pt antenna (see solid lines in Fig.~\ref{fig2}(a) and Fig.~\ref{fig2}(b)).
Switching the setup to the DC regime shows that the DC pulse is also detected with practically the same delay (see Fig.~\ref{fig2}(c) and Fig.~\ref{fig2}(d)). As one can see from the Figure the reversal of YIG magnetization results in a change of the detected DC voltage sign proving its ISHE nature.~\cite{saitoh-2006} This experiments directly demonstrates the detection of magnon spin transport by the inverse spin Hall effect.

An increase in the duration $\tau_\mathrm{s}$ results in an increase in the amplitude and the duration of both the AC and DC signals,~\cite{comment-2} additionally pointing to their spin-wave origin.
However, a more careful comparison of the SW intensity and the ISHE voltage shows that the maximum of the DC signal is slightly shifted in time compared to the AC signal. Furthermore, the ISHE pulse shows a slower decay. These effects have been recently reported in Ref.~\cite{jungfleisch-APL} and are due to the contribution to the spin pumping of secondary spin waves. In the linear case ($P_\mathrm{s} = 10$~mW) these waves are known to be dipolar-exchange spin waves (DESWs) excited as a result of elastic two-magnon scattering of the propagating wave.\cite{Gurevich, Sparks, melkov-2004} Due to their small wavelengths, DESWs are not detectable by the inductive antenna (Pt strip in our case),\cite{YIG-magnonics, sandweg-PRL} but they contribute effectively to the spin pumping and the ISHE voltage.\cite{jungfleisch-APL} Besides, due to their smaller velocity and lower damping, DESWs produce the ISHE voltage even after the traveling magnons have left the Pt region. This results in the slower decay of the DC pulse.

We have estimated the contribution of the DESWs to the ISHE voltage pulse. Since the antenna's width is larger than the DESWs propagation length, but much smaller than the decay distance of primary SW, one can neglect propagation of DESWs and feedback reaction of DESWs on the primary SW dynamics. In this case the dynamics of the number $n_k$ of magnons in $k$-th DESW mode can be modeled as
\begin{equation}\label{eq1}
{\partial n_k}/{\partial t} + 2 \Gamma_k n_k = 2 R_k n_0(t),
\end{equation}
where $\Gamma_k$ is the damping parameter of the $k$-th DESW, $R_k$ is the intensity of 2-magnon scattering from SW to $k$-th DESW, and $n_0(t)$ is the time profile of primary SW under the antenna. Assuming, for simplicity, that all DESWs have the same damping rate $\Gamma_k = \Gamma = const$, one can derive simple expression for the total density $n_\mathrm{sum}(t) = n_0(t) + \sum_{k}{n_k(t)}$ of magnons under the antenna:
\begin{equation}\label{eq2}
n_\mathrm{sum} = n_0(t) + 2 R \int_{-\infty}^{t}{n_0(t) e^{-2 \Gamma (t-t') dt'}},
\end{equation}
where $R = \sum_k{R_k}$ is the total intensity of 2-magnon scattering.

The numerically calculated normalized density $n_\mathrm{sum}$ is shown in Fig.~\ref{fig2}(a) and Fig.~\ref{fig2}(b) with dashed lines. The damping $\Gamma = 7\cdot10^6$~rad/s and the scattering efficiency $R = 20\cdot10^6$~rad/s were used as fitting parameters.\cite{comment-3} Accumulation of DESW during the duration of the primary SW pulse results in a significant DESWs contribution to the ISHE voltage: at the maxima, $n_\mathrm{sum}/n_0 \approx$~1.8 for $\tau_\mathrm{s} = 50$~ns and $n_\mathrm{sum}/n_0 \approx$~2.1 for $\tau_\mathrm{s} = 100$~ns. As a result one sees in Fig.~\ref{fig2} the slower decay and the shift of the maxima for $n_\mathrm{sum}$.

\begin{figure}[t]
\includegraphics[width=0.95\columnwidth]{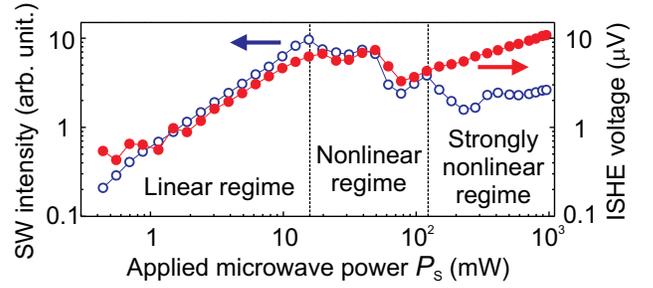}
\caption{\label{fig3} (Color online) Transmitted SW intensity (open circles) and ISHE voltage (filled circles) as
functions of the applied microwave power $P_\mathrm{s}$. The pulse duration is $\tau_\mathrm{s} = 100$~ns.}
\end{figure}

The dependencies of the SW intensity and ISHE voltage on the applied microwave power $P_\mathrm{s}$ are shown in Fig.~\ref{fig3}. Three different regions can be assigned as it is shown in the figure. In the first region the detected ISHE voltage is practically proportional to the SW intensity. In the second region, with increasing power, the SW intensity as well as the ISHE voltage decrease. This drop is due to the onset of nonlinear multi-magnon scattering processes which partially suppress the propagating SW packet.\cite{Sparks, Lvov} The suppressing takes place mostly under the input antenna and the linear (but already decreased in intensity) traveling waves propagate toward the Pt strip. The secondary magnons, which are excited near the input antenna, are rather slow and practically do not reach the detector area. Nevertheless, a further increase of $P_\mathrm{s}$ brings the system to a new regime: the density of the secondary magnons is so high that even a small percentage of them, that reaches the Pt detector, contributes to the ISHE voltage more than the suppressed traveling SWs. This effect is visible as an increase of the ISHE voltage independent from the detected SW intensity.

\begin{figure}[t]
\includegraphics[width=0.95\columnwidth]{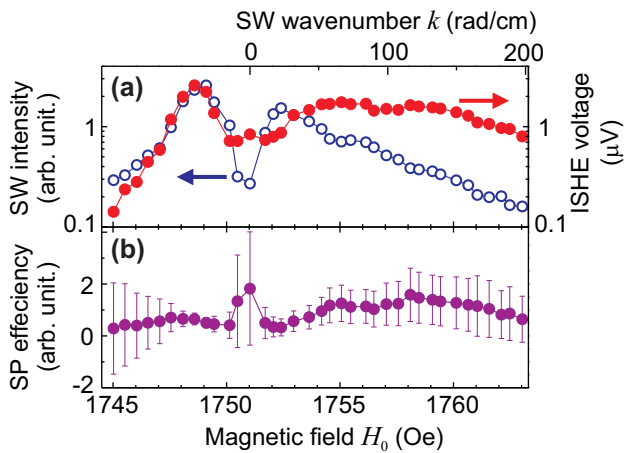}
\caption{\label{fig4} (Color online) The transmitted SW intensity (open circles) and ISHE voltage (filled circles) are shown in panel (a) as functions of the bias magnetic field $H_0$. Corresponding to the $H_0$ BVMSW wavenumbers $k$ are indicated on the top scale. The spin pumping efficiency in panel (b) is found as a ratio of the ISHE voltage to the SW intensity with additional account of the SW detection efficiency $F(k)$. The carrier signal frequency $f_\mathrm{s}=7$~GHz, the signal duration $\tau_\mathrm{s}=1$~$\mu$s.}
\end{figure}

In the experiment we use a narrow input antenna which allows for excitation of traveling spin waves in a certain range of wavenumbers $k$ by varying the field $H_0$. The transmitted SW intensity as a function of $H_0$ is shown in Fig.~\ref{fig4}(a) with open circles. This dependence is standard for BVMSW: it has the maximum slightly above the point of uniform precession $k = 0$ and decreases with increase in $k$ due to the drop of excitation and AC detection efficiencies.~\cite{YIG-magnonics} The dependence of the ISHE voltage on $k$ is shown in the same figure with filled circles. One can see that with an increase in $H_0$ the voltage decreases slower in comparison to the SW intensity.
This might suggest that the spin pumping efficiency, defined as the ratio of the detected ISHE voltage to the SW intensity, increases with an increase in $k$.
However, this behavior can be easily explained by taking into account $k$-dependence of the AC detection efficiency $F(k) = (\sin(kw/2)/(kw))^2$ by the Pt antenna ($w = 200$~$\mu$m is the width of the Pt strip).\cite{ganguly-1975} As one can see from Fig.~\ref{fig4}(b) the corrected spin pumping efficiency is independent of the $k$-vector of the travelling spin wave within the experimental uncertainty.~\cite{comment-5}

In conclusion, we have detected a magnon spin transport over a macroscopic distance as spin and charge currents in the non-magnetic metal attached to the ferrimagnetic spin-wave waveguide. In addition, the contribution of the secondary excited magnons to the ISHE voltage is measured and estimated theoretically. It has been shown that the contribution of the secondary magnons to the ISHE voltage is comparable to that of the originally excited traveling magnons and leads to delay and shape distortion of the ISHE voltage pulse. The field dependent measurements have shown that the spin pumping efficiency in YIG-Pt bi-layers does not depend on the spin-wave wavelength.

We thank G.~E.~W.~Bauer and G.~A.~Melkov for valuable discussions and the Nano-Structuring Center, TU Kaiserslautern, for technical support. This work was supported in part by the Grant No. ECCS-1001815 from National Science Foundation of the USA.


\begin{thebibliography}{sotief}
\bibitem{comment-1} In order to ensure the quality of the YIG/Pt interface we tested the YIG-Pt sample using the ferromagnetic resonance technique.\cite{jungfleisch-APL} The ISHE voltage reached values up to $\pm 1.5$~mV for an applied microwave power of 1~W.
\bibitem{comment-3} This value of $\Gamma$ corresponds to resonance linewidth $\Delta H = 0.8$~Oe which is larger than the literature values of 0.1-0.2 Oe. This increase of the DESW damping under the Pt antenna might be caused by the additional DESW losses due to the  spin-pumping mechanism.
\bibitem{comment-5} The region left to the point $k = 0$ corresponds to the magnetostatic surface spin-wave mode propagating under some angle to the bias magnetic field. No dependence of spin-pumping efficiency on the SW wavenumbers is visible here as well.
\end{thebibliography}
\end{document}